\documentclass{elsart}

\usepackage{natbib}

\usepackage{amssymb}
\usepackage{graphicx}        

\begin{document}

\begin{frontmatter}



\title{Stars and Gas in the Inner Parts of Galaxies seen in SAURON Integral Field
Observations}


\author[Kap]{R.F. Peletier\ead{peletier@astro.rug.nl}},
\author[Kap]{K. Fathi\thanksref{cor1}}

\address[Kap]{Kapteyn Astronomical Institute, University of Groningen, The
Netherlands}

\corauth[cor1]{Current address: Instituto de Astrofisica de Canarias,
Tenerife, Spain}

\author[Hat]{E.L. Allard},
\author[Hat]{J.H. Knapen}, 
\author[Hat]{M. Sarzi}
\address[Hat]{Centre for Astrophysics Research, University of Hertfordshire, UK}

\author[Lei]{G. van de Ven\thanksref{cor2}},
\author[Lei]{J. Falcon-Barroso\corauthref{cor3}}, 
\author[Lei]{M. Cappellari\corauthref{cor4}}, 
\author[Lei]{P.T. de Zeeuw}

\address[Lei]{Sterrewacht Leiden, Leiden, the Netherlands}
\corauth[cor2]{Current address: Institute for Advanced Study, Princeton, NJ,
USA}
\corauth[cor3]{Current address: Estec, Noordwijk, the Netherlands}
\corauth[cor4]{Current address: Sub-Department of Astrophysics, University of
Oxford, UK}

\author[Lyo]{E. Emsellem}
\address[Lyo]{CRAL - Observatoire de Lyon, Saint Genis Laval, France}

\begin{abstract}
We give two examples of spiral galaxies that show non-circular gas motions 
in the inner kiloparsecs, from SAURON integral field spectroscopy. We use 
harmonic decomposition of the velocity field of the ionized gas to study the
underlying mass distribution, employing linear theory. The higher order harmonic
terms and the main kinematic features of the observed data are consistent with
an analytically constructed simple bar model. We also present maps of a number
of strong absorption lines in M 100, derive simple stellar populations and
correlate them with features in the gas kinematics. 
\end{abstract}


\end{frontmatter}

\section{Introduction}
\label{Intro}

Although circular rotation is the dominant kinematic feature of galactic disks,
it is clear from observations and theory that in many galaxies non-circular
motions are also present (e.g. Freeman et al. 1965, Shlosman et al. 1989).
Non-axisymmetric motion can, in fact, be a crucial ingredient in determining
the evolution of a galaxy, by triggering star formation, or by building a
central mass concentration.  

We know that many galaxies show non-axisymmetric mass-distributions:
near-infrared photometric studies show that at least 70\% of the galaxies in the
nearby Universe show strong or weak bars (Eskridge et al. 2000, Knapen et al.
2000a). Such asymmetric features in the mass-distribution will produce
non-circular motions, although also other effects, such as 
interactions with other galaxies, can cause them. Gas, being more responsive to perturbations
and instabilities than stars, will show these non-circular motions more easily. 
Athanassoula (1992) shows how gas streaming manifests itself 
in galaxies with strong bars. Her calculations show that one
can use the motion of the gas, as measured in two dimensions, to
derive information about the mass distribution in galaxies, especially if these
are relatively isolated. This has not been done very often in the literature up
to now, since good, high resolution data is scarce, and up to recently a good
theoretical framework was lacking.

Recently, Wong, Blitz \& Bosma (2004) have quantified  the strength of bars and
spiral arms using HI and CO velocity fields of a number of nearby galaxies.
They used a harmonic decomposition formalism developed by Schoenmakers et al.
(1997), and compared the results with
analytical  models. Here we do the same with much higher resolution optical
emission line data, obtained with the Integral Field Spectrograph SAURON, for
NGC~5448, an early-type spiral galaxy. More details can be found in Fathi et al. (2005;
{\bf F05}).  

From the SAURON data one can not just measure emission lines, but also
absorption lines. In the second part of this paper we present absorption line
maps of the central regions of M100, together with some simple stellar
population models, recently published by Allard et al. (2006; {\bf A06}). Absorption lines,
contrary to integrated colours, are relatively insensitive to dust absorption,
so that more accurate ages and other stellar population parameters can be
obtained.

In the future we plan to perform a harmonic decomposition (also called
kinemetry, see Krajnovic et al. 2006) on the full sample SAURON early-type
spirals, presented in Falc\'on-Barroso et al. (2006), and analyse statistically
the non-axisymmetric motions in the centers of nearby spirals.

\section{A bar signature in the gas velocity field of NGC 5448}
\label{bar}

In Fig. 1 we show the observed velocity field of NGC 5448 from the [OIII]
emission line at 5007\AA. Clearly visible is
the streaming at the bottom right and top left. Decomposing the velocity field
into Fourier harmonic components using

\begin{equation}
\label{eq:expansion}
V_\mathrm{los}= V_\mathrm{sys} + 
		\sum^{k}_{n=1}{\big(c_n(R)\cos\,n\psi + s_n(R)\sin\,n\psi\big)\sin i}, 
\end{equation}

where $k$=3, c$_n$ and s$_n$ give us information about the nature of the
perturbations. As mentioned in F05, c$_1$, c$_3$ and s$_3$ give
information about possible $m$ = 2 perturbations. These coefficients are 
plotted in the figure, along with an analytic bar model, which shows how well such a model can
reproduce the complex observations (for details see 
F05). At the bottom left we show the normalized s$_1$ vs. s$_3$ term. A negative
slope in this diagram shows that the velocity field indicates the presence of a
bar, and not a warped disk. 

Another galaxy with strong non-axisymmetric gas motions in the central regions
is M 100 (NGC 4321). Knapen et al. (2000b) present an H$\alpha$ image with
strong streaming motions, and model it with a numerical model. Wong et al.
(2004) perform a harmonic Fourier decomposition. In Allard et al. (2006) we
show that the [OIII] map of M100 accurately reproduces the non-axisymmetric
features, and that these features are also found in the stellar kinematics.

These two cases show that the SAURON data are well suited to measure bars,
purely on the basis of the kinematics. The kinematics offer an easier way, as
opposed to imaging, to measure the bar potential, and give usable results even
in the presence of considerable dust extinction.

\begin{figure}[t]
\centering
\includegraphics[width=14cm]{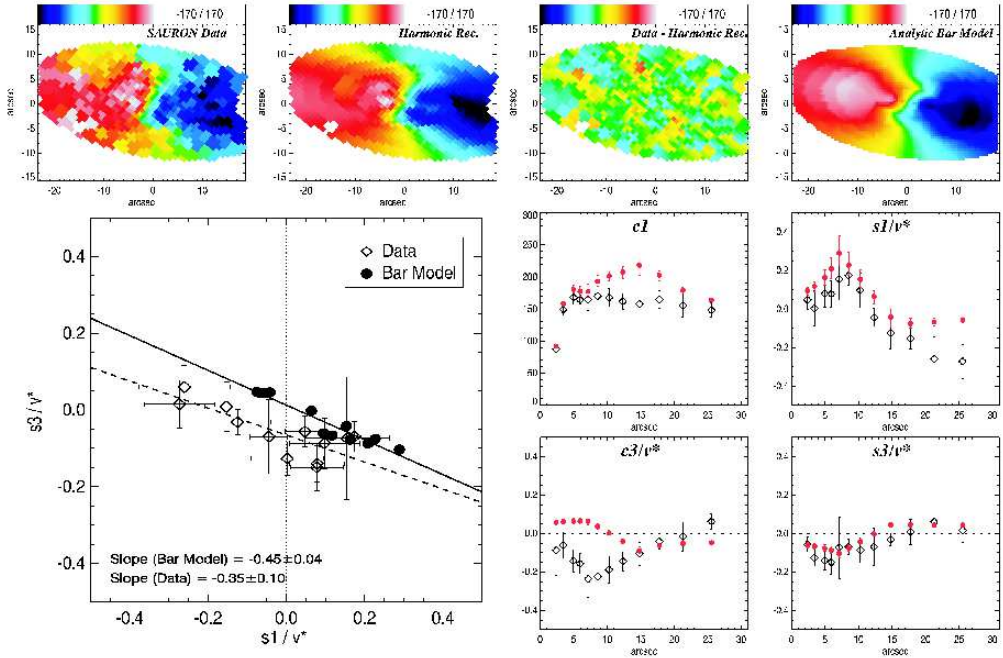}
\caption{Top row, from left to right: observed SAURON ([OIII])
gas velocity field of NGC~5448; reconstruction based on the
harmonic decomposition;
residual field (data - harmonic reconstruction); and the analytic
bar model, which best reproduces the main kinematic features of
the observed gas velocity field. Middle and bottom row: (right) the
harmonic parameters as a function of radius of each ring,
where $v^* =c_1 \sin i$. The over-plotted red filled circles are the
analytically calculated first and third harmonic terms for the bar
model (the second terms are zero by construction). (left): first vs. 
third sinusoidal harmonic term, showing the applicability of our bar
model (from F05).
}
\label{fig:peletier_fig1}
\end{figure}

\section{Stellar populations in the inner regions of M 100}
\label{m100}

\begin{figure}[t]
\centering
\includegraphics[width=14cm]{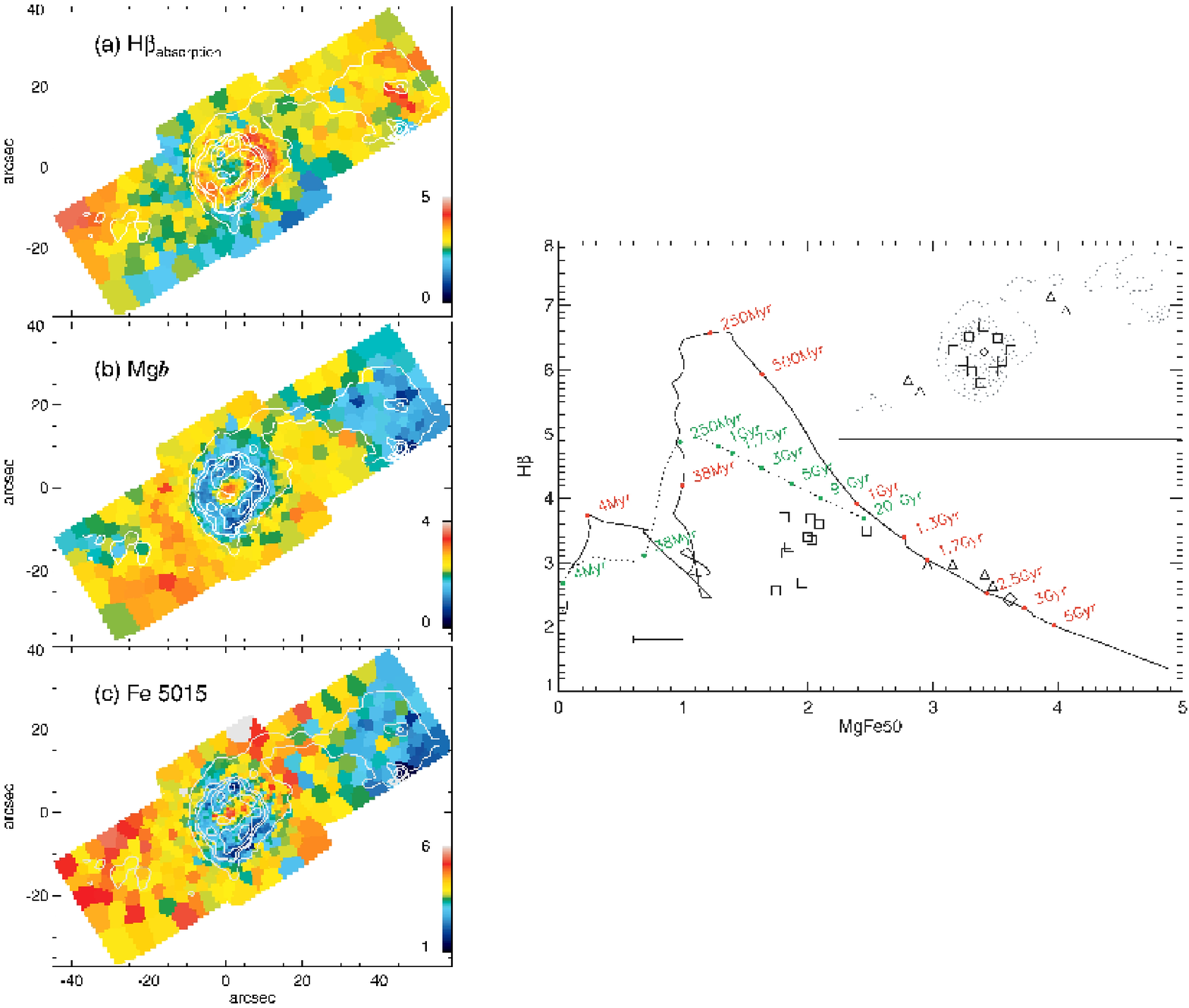}
\caption{(Left:) Absorption line maps of (a) H$\beta$, (b) Mg~b and (c) Fe 5015.
(Right:) H$\beta$ against MgFe50 in the central regions of
M100 for several apertures  (see inset for their location, relative to the H$\beta$ emission 
map). The solid line shows the track for an SSP starburst model, with indicative
ages shown in red. The dashed line shows the track for a continuous starburst
model with a constant star formation rate. From {\bf A06}.}
\label{fig:peletier_fig2}
\end{figure}

SAURON not only produces maps of emission lines, due to ionized gas, but also
absorption line maps for some strong lines. In Fig.~2 we show H$\beta$, Mg~b
and Fe 5015 lines, derived in the Lick system (Worthey et al. 1994). To obtain these
maps, one has to remove the emission lines first (see Sarzi et al. 2006). 
The maps show that the center of M 100 is
relatively old. Going outwards, a very young ring follows, after which there is
a region with stars of about 10$^8$ years (see {\bf A06}). On the right side of Fig.~2 we show
several regions in and around the ring in a plot of MgFe50 (=
$\sqrt {( {\rm Mg~b} \times\ {\rm Fe 5015}}$) against H$\beta$, together with SSP models of
Bruzual \& Charlot (2003). In general, the models are able to fit the data. In
some places in the star forming ring, however, at MgFe50 $<$ 2\AA\, SSPs are
not able to explain the observations any more. Here a solution would be a mix
of a 1.5 Gyr old stellar population and one of more younger bursts. 
Such a picture ties in extremely well
with the theoretical and numerical understanding we have of the nuclear
ring zone in M100 (Knapen et al. 1995).
When one looks in detail, the youngest stellar
populations are found at the contact points between the ring and the dust
lanes. More detailed modelling of the stellar populations in this galaxy is
done in {\bf A06}.

Another indication of how complex the stellar populations and the distribution
of the
interstellar matter in the inner regions of M 100 are, is shown in Fig.~3. Here
we show {\it Spitzer Space Telescope (SST)} images at 4 different wavelengths. At the
shorter wavelengths (3.6 and 4.5$\mu$m) the maps look like the 2.2$\mu$m map of
Knapen et al. (1995), showing the nuclear part of the bar. These maps are relatively free of
dust extinction. At larger wavelengths (5.8 and 8$\mu$m) the emission comes
mostly from tiny dust grains and polycyclic aromatic hydrocarbons (PAHs). These
latter maps show especially the star formation regions, similar to the $B$-band
image of Knapen et al. (1995). A comparison of the exponential scale lengths of
bulge and disk between several optical wavelengths and the 3.6 $\mu$m gives
almost identical values, reinforcing the conclusion by Beckman et al. (1996)
that the galaxy does not contain much dust, except in localised areas.

\begin{figure}[t]
\centering
\includegraphics[width=10cm]{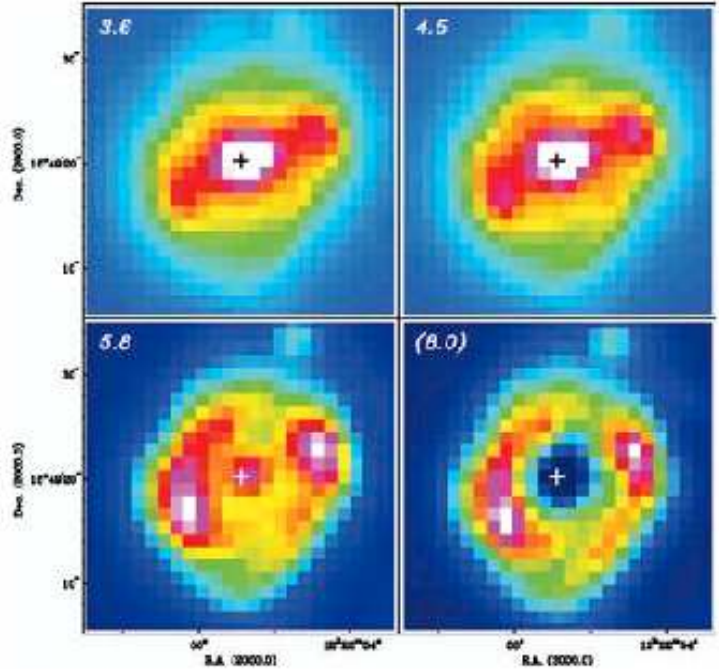}
\caption{{\it SST} IRAC 3.6, 4.5, and 5.8 $\mu$m images of the central regions of
M100, as well as the dust only 8$\mu$m image (see {\bf A06}). The nucleus of the
galaxy is indicated with a cross.}
\label{fig:peletier_fig4}
\end{figure}

\section{Discussion and Conclusions}
\label{Discussion}

In the previous sections we have shown two applications of integral field
spectroscopy aimed at understanding the nature of the central regions of spiral
galaxies. We show that, in contrast to long slit spectroscopy, we can detect
streaming motions of the ionized gas, and use them to measure the potential.
Although something similar can also be done from the photometry, there are more
assumptions involved here, since one has to look through the dust, and make
assumptions for {\it M/L} ratios as a fuction of position. In a galaxy like M100,
where the stellar populations are strongly varying, estimating M/L purely from
colours and absorption line strengths will not be easy at all. Up to now we have
only analyzed the Sa galaxy NGC 5448 in detail. Once we have analyzed the full
sample of Falc\'on-Barroso et al. (2006) we will be able to derive statistics on
the number of bars, and their strengths, based for the first time on kinematic
data. These statistics will be important, among other topics, in assessing whether AGN activity is
related to the presence of bars (e.g., Shlosman et al. 1989).

In the second application we show line strength maps of a galaxy with a
circumnuclear ring. Since most Sb galaxies are very dusty in their inner
regions one cannot learn too much from broad band colours, unless the galaxy
has a special orientation (e.g., Peletier et al. 1999). From three line
strength maps we find that the galaxy has zones with different ages. In the
ring itself we show that apart from the young stars there is also an underlying
old stellar population.  M 100 is a beautiful case where many phenomena can be
seen that play a role in the inner regions of spiral galaxies. The outer
and inner parts of its bar have been detected both in near-IR (ground-based and Spitzer)
photometry and in the gas kinematics, and a ring separating the two. Abundant
star formation occurs within and near the ring, with 2 hot spots at the end of the inner
bar. Young stars are formed where the velocity dispersion of the gas is low
(Allard et al. 2005). At present we are studying the stellar populations of the
sample of Sa galaxies of Falc\'on-Barroso et al. (2006) in the same way. Many
galaxies of that sample show young stellar populations in circumnuclear rings
or disks, showing that for most galaxies the central regions are not dominated
by a bright, old stellar bulge. More details can be found in Peletier et al.
(in preparation).





\begin{thebibliography}{}

\bibitem[Allard et al.(2005)]{2005ApJ...633L..25A} Allard, E.~L., Peletier, 
R.~F., \& Knapen, J.~H.\ 2005, ApJl, 633, L25 
\bibitem[Allard et al.(2006)]{2006MNRAS.tmp..867A} Allard, E.~L., Knapen, 
J.~H., Peletier, R.~F., \& Sarzi, M.\ 2006, MNRAS, 867 
\bibitem[Athanassoula(1992)]{1992MNRAS.259..345A} Athanassoula, E.\ 1992, 
MNRAS, 259, 345 
\bibitem[Beckman et al.(1996)]{1996ApJ...467..175B} Beckman, J.~E., 
Peletier, R.~F., Knapen, J.~H., Corradi, R.~L.~M., \& Gentet, L.~J.\ 1996, 
ApJ, 467, 175 
\bibitem[Bruzual \& Charlot(2003)]{2003MNRAS.344.1000B} Bruzual, G., \& 
Charlot, S.\ 2003, MNRAS, 344, 1000 
\bibitem[Eskridge et al.(2000)]{2000AJ....119..536E} Eskridge, P.~B., et 
al.\ 2000, AJ, 119, 536 
\bibitem[Falc{\'o}n-Barroso et al.(2006)]{2006MNRAS.369..529F} 
Falc{\'o}n-Barroso, J., et al.\ 2006, MNRAS, 369, 529 
\bibitem[Fathi et al.(2005)]{2005MNRAS.364..773F} Fathi, K., van de Ven, 
G., Peletier, R.~F., Emsellem, E., Falc{\'o}n-Barroso, J., Cappellari, M., 
\& de Zeeuw, T.\ 2005, MNRAS, 364, 773 
\bibitem[Freeman(1965)]{1965MNRAS.130..183F} Freeman, K.~C.\ 1965, MNRAS, 
130, 183 
\bibitem[Knapen et al.(1995)]{1995ApJ...443L..73K} Knapen, J.~H., Beckman, 
J.~E., Shlosman, I., Peletier, R.~F., Heller, C.~H., \& de Jong, R.~S.\ 
1995, ApJl, 443, L73 
\bibitem[Knapen et al.(2000)]{2000ApJ...529...93K} Knapen, J.~H., Shlosman, 
I., \& Peletier, R.~F.\ 2000a, ApJ, 529, 93 
\bibitem[Knapen et al.(2000)]{2000ApJ...528..219K} Knapen, J.~H., Shlosman, 
I., Heller, C.~H., Rand, R.~J., Beckman, J.~E., \& Rozas, M.\ 2000b, ApJ, 
528, 219 
\bibitem[Krajnovi{\'c} et al.(2006)]{2006MNRAS.366..787K} Krajnovi{\'c}, 
D., Cappellari, M., de Zeeuw, P.~T., \& Copin, Y.\ 2006, MNRAS, 366, 787 
\bibitem[Peletier et al.(1999)]{1999MNRAS.310..703P} Peletier, R.~F., 
Balcells, M., Davies, R.~L., Andredakis, Y., Vazdekis, A., Burkert, A., \& 
Prada, F.\ 1999, MNRAS, 310, 703 
\bibitem[Sarzi et al.(2006)]{2006MNRAS.366.1151S} Sarzi, M., et al.\ 2006, 
MNRAS, 366, 1151 
\bibitem[Schoenmakers et al.(1997)]{1997MNRAS.292..349S} Schoenmakers, 
R.~H.~M., Franx, M., \& de Zeeuw, P.~T.\ 1997, MNRAS, 292, 349 
\bibitem[Shlosman et al.(1989)]{1989Natur.338...45S} Shlosman, I., Frank, 
J., \& Begelman, M.~C.\ 1989, Nature, 338, 45 
\bibitem[Wong et al.(2004)]{2004ApJ...605..183W} Wong, T., Blitz, L., \& 
Bosma, A.\ 2004, ApJ, 605, 183 
\bibitem[Worthey et al.(1994)]{1994ApJS...94..687W} Worthey, G., Faber, 
S.~M., Gonzalez, J.~J., \& Burstein, D.\ 1994, ApJs, 94, 687 



\bibitem[]{}

\end{thebibliography}
\end{document}